\documentclass[12pt]{JHEP3}
\usepackage{multicol}
\usepackage{amsmath,amsfonts,amssymb,amsthm}

\input epsf
\input xy
    \xyoption{matrix}
    \xyoption{arrow}

\def\M{\mathcal{M}}

\def\GL{\operatorname{GL}}
\def\diag{\operatorname{diag}}
\def\Der{\operatorname{Der}}

\def\H{\mathcal{H}}

\def\D{\mathcal{D}}
\def\A{\mathfrak{A}}
\def\a{\mathfrak{a}}
\def\g{\mathfrak{g}}
\def\h{\mathfrak{h}}
\def\gl{\mathfrak{gl}}
\def\R{\mathbb{R}}
\def\Z{\mathbb{Z}}
\def\C{\mathbb{C}}
\def\F{\mathcal{F}}

\def\To{\longrightarrow}
\def\d{\partial}
\def\eps{\epsilon}
\def\la{\lambda}

\def\<{\left\langle}
\def\>{\right\rangle}

\def\fZ{\mathfrak{Z}}
\def\vp{\varphi}
\def\Tr{\operatorname{Tr}}
\def\Aut{\operatorname{Aut}}

\def\ox{\otimes}

\def\te#1{\text{#1}}
\def\norm#1{\left\| #1 \right\|}

\def\proofthm#1{\vskip 0.05 in
        \noindent {\it Proof of Theorem \ref{#1}.} $\ \ $ }

\def\ritem#1{\item[{\rm #1}]}
\def\f#1#2{\frac{#1}{#2}}

\def\d{\partial}
\def\half{\frac{1}{2}}

\newtheorem{corollary}{Corollary}
\newtheorem{theorem}{Theorem}
\newtheorem{lemma}{Lemma}
\theoremstyle{definition}
\newtheorem{definition}{Definition}
\theoremstyle{remark}

\newtheorem{example}{Example}

\def\be{\begin{equation}}
\def\ee{\end{equation}}

\title{Quantum Field Theory and the Space of All Lie
Algebras}

\author{{ William Gordon Ritter}\\
{\small Harvard University Department of Physics} \\
{\small 17 Oxford St., Cambridge, MA 02138} \\
{\it Email: ritter@fas.harvard.edu}}

\abstract{ The space $\M_n$ of all isomorphism classes of
$n$-dimensional {Lie} algebras over a field $k$ has a natural
non-Hausdorff topology, induced from the Segal topology by the
action of $\GL(n)$. One way of studying this complicated space is
by topological invariants. In this article we propose a new class
of invariants coming from quantum field theory, valid in any
dimension, inspired by Jaffe's study of generalizations of the
Witten index.}


\preprint{math-ph/0304031}

\begin{document}

\vskip 0.1 in
\section{The Space of All Lie Algebras}  \label{sec:deformations}

Let $\g$ be a finite-dimensional Lie algebra over an arbitrary
field $k$. In what follows we will mostly assume $k$ is $\R$ or
$\C$.

\begin{definition} \label{def:structensor}
The \emph{structural tensor} $f$ is defined to be the element of
$\g^* \otimes \g^* \otimes \g$ given by considering the bracket as
a skew-symmetric bilinear mapping $\g \otimes \g \to \g$ (which is
the same as a map $\wedge^2 \g \to \g$).
\end{definition}

In a basis $\{E_a\}$ of $\g$, the components of the structural
tensor satisfy $[E_a, E_b] = {f_{ab}}^c E_c$. The statement that
$\g$ forms a Lie algebra is equivalent to the following relations
for the structure constants:
\begin{equation} \label{eq:str-const}
{f_{ij}}^k + {f_{ji}}^k = 0, \quad {f_{ij}}^\ell \cdot {f_{\ell
k}}^m + {f_{jk}}^\ell \cdot {f_{\ell i}}^m + {f_{ki}}^\ell \cdot
{f_{\ell j}}^m = 0
\end{equation}
These relations are more compactly written as
\[
{f_{(ij)}}^k = {f_{(ij}}^l {f_{k) l}}^m = 0
\]

The space of all structure constants of $n$-dimensional Lie
algebras inherits a topology from $k^{n^3}$. We call this the
\emph{Segal topology} because of its relevance to Ref.~\cite{Se}.
The space of all sets $\{ {f_{ij}}^k \}$ satisfying
\eqref{eq:str-const} is a subvariety $W^n \subset k^{n^3}$ of
dimension
\[
\dim W^n \leq n^3 - \frac{n^2(n+1)}{2}  = \frac{n^2(n-1)}{2}.
\]

The space $\M_n$ of all isomorphism classes of $n$-dimensional
{Lie} algebras over $k$ has a natural weakly separating (i.e.
$T_0$, not $T_1$) non-{Hausdorff} topology $\kappa^n$, induced
from the Segal topology by the action of $\GL(n)$. Basis
transformations induce $\GL(n)$ tensor transformations between
equivalent structure constants,
\begin{equation}
C^k_{ij} \sim (A^{-1})^k_h\ C^h_{fg}\ A^f_i\ A^g_j, \ \ A \in
\GL(n)\, .
\end{equation}
One can either define $\M_n = W^n/\GL(n)$, or avoid the
basis-dependent notation entirely, and simply define $\M_n$ as the
set of all $n$-dimensional structural tensors in the sense of
Definition \ref{def:structensor}.

For $n=3$ the structure constants can be written as
\be \label{n=3}
{f_{ij}}^k = \eps_{ijl}(n^{lk}+\eps^{lkm}a_m),
\ee
where $n^{ij}$ is symmetric and $\eps_{ijk} = \eps^{ijk}$ totally
antisymmetric with $\eps_{123}=1$. With \eqref{n=3} the
constraints are equivalent to $n^{lm}a_m = 0,$ which are $3$
independent relations. There is a classification due to Behr of
Lie algebras in $\M_3$ according to possible inequivalent
eigenvalues of $n^{lm}$ and values of $a_m$ (see \cite{Lan}).

Invariants of real {Lie} algebras have been calculated for $n\leq
5$ by {Patera, Sharp, Winternitz} and {Zassenhaus} \cite{PaSWZ}.
In this article we propose a new class of invariants, valid in any
dimension, inspired by Jaffe's study of generalizations of the
Witten index \cite{QHA,QI}.

\section{Deformation Theory}

\begin{definition}
A \emph{Lie algebra deformation} (or simply, a deformation) is a continuous
curve $f : [0, \eps] \to \M_n$, where $\eps > 0$. The deformation
is said to be \emph{trivial} if all $\g(t)$ are isomorphic, where
$\g(t)$ is the Lie algebra with structural tensor $f(t)$.
\end{definition}

We denote by $\Der \g$ the set of all linear maps $D : \g \to \g$
such that $D[a,b] = [Da,b] + [a,Db]$.

Let $G$ and $H$ be Lie groups and let $B$ be an action of $G$ on
$H$ by group homomorphisms with $B:G \times H \to H$ smooth. Then
the semidirect product group $G \ltimes H$ is a Lie group with
group operation: $(g_1, h_1) (g_2, h_2) = (g_1g_2, h_1
B(g_1)h_2)$. We denote by $\beta : G \to \Aut\h$ the map defined
by $\beta(g) = T_e B(g)$ and $b : \g \to \Der\h$ the differential
of $\beta$ at the identity. Then $b$ defines a semidirect product
of Lie algebras $\g \ltimes_b \h$ by
\[
[(X_1,Y_1), (X_2,Y_2)] = ([X_1,X_2], [Y_1,Y_2] + b(X_1)Y_2 -
b(X_2)Y_1)
\]
and this is the Lie algebra of $G \ltimes H$.

\begin{example}  \label{ex:nonisomorphic}
An $\R$-linear map $\vp : \R \to \gl_2(\R)$ is determined by its
value at 1. Let $\a(t) = \R \ltimes_{\vp_t} \R^2$, where $\vp_t(1)
= \diag\big( (1+t)\lambda,\, (1+t+\alpha t^2)\mu \big)$ for
$\lambda, \mu \ne 0$ and $\alpha > 0$. Then $t \to \a(t)$ is
continuous and the Lie algebras $\a(t)$ are pairwise
nonisomorphic.
\end{example}

As a vector space, $\a(t) \cong \R^3$, so we can choose the
standard basis $e_1, e_2, e_3$. With this choice of basis, the
commutation relations of Example \ref{ex:nonisomorphic} are
\begin{equation} \label{ni2}
[e_1, e_2] = (1+t)\lambda e_2, \quad [e_1, e_3] = (1 + t + \alpha
t^2) \mu e_3, \quad [e_2, e_3] = 0
\end{equation}
In terms of the structure constants, $f_{1 j}^k = a(j) \delta_j^k$
for $a(2) = (1+t)\lambda, a(3) = (1 + t + \alpha t^2) \mu$, and
$f_{2j}^k = -\delta_j^1 \delta_2^k a(2)$.

For any finite-dimensional Lie algebra $\g$, there exists
a unique (up to isomorphism) connected, simply-connected Lie
group $G$ having $\g$ as its tangent algebra.
It follows that the classification of simply-connected Lie groups
reduces to the classification of the corresponding Lie algebras.
In particular, Example \ref{ex:nonisomorphic} proves the existence
of pairwise nonisomorphic continuous deformations of Lie groups.

\section{Lie Algebra Cohomology and BRST Theory}

We review the connection between
Lie algebra cohomology and quantization of gauge theories, setting notation for
later sections.

Let $\{e_\alpha\}$ be a basis of $V$ and
$\{T_i\}$ of $\mathfrak{g}$, with ${f_{ij}}^{k}$ the structure
constants, so that $[T_i,T_j] = {f_{ij}}^{k} T_k$. Let
$t_i$ be the generators in the representation $\rho$, i.e.
$t_i = \rho(T_i)$. A $V$-valued $n$-\textit{cochain} is an
antisymmetric linear map $u : \g^n \rightarrow V$ specified by
\begin{equation}
\label{cochain}
u(T_{i_1}, \ldots, T_{i_n}) = u_{i_1\ldots i_n}^{\alpha}e_\alpha,
\qquad\qquad (n \geq 0)
\end{equation}
with coefficients $u_{i\ldots j}^{\alpha}$ being totally antisymmetric
in lower indices. The \textit{coboundary operator} $\delta$
increases the valence of arbitrary $n$-cochain $u$ by one, sending it
to $(n+1)$-cochain $\delta u$. Explicitly for $x_k \in \g$, we have
\begin{multline}
\label{cobound}
\delta u\,(x_1,\,\ldots\,,x_{n+1}) \doteq
\sum_{k=1}^{n+1}(-)^{k+1}\rho(x_k)u(x_1,\,\ldots\,,{\not\!x}_k,\,
\ldots\,,x_{n+1}) \\
+ \sum_{j<k}(-)^{j+k}
u([x_j,x_k],x_1,\,\ldots\,,{\not\!x}_j,\,\ldots\,,{\not\!x}_k,
\,\ldots\,,x_{n+1})\,.
\end{multline}
In component form,
\begin{equation}
\label{}
(\delta u)_{i_1\ldots i_{n+1}}^{\alpha}
= \sum_{k=1}^{n+1}(-)^{k+1}\,(t_{i_k})_{\beta}^{\alpha}
u_{i_1\ldots\not i_k\ldots i_{n+1}}^{\beta}
 + \sum_{j<k}(-)^{j+k}f_{i_j i_k}^{m}
u_{m i_1\ldots\not i_j\ldots\not i_k\ldots i_{n+1}}^{\alpha}\,.
\end{equation}
The coboundary operator is nilpotent, $\delta\circ\delta = 0$. The
corresponding cohomology groups are denoted
by $H^n(\g,V)$. The exterior derivative
provides one example of such a coboundary operator (with
$\g$ being the Lie algebra $\mathcal{X}(M)$ of vector fields,
and $V$ the algebra of functions on a manifold $M$)\,.

Due to the antisymmetry of (\ref{cochain}), $n$-cochains may be
considered as elements of
$\wedge^n \mathfrak{g}^*\otimes V$. A particularly convenient
notation for cochains is to view the basis elements $\{c^i\}$ of
$\mathfrak{g}^*$ as odd (anticommuting) objects:
$c^i\wedge c^j = c^i c^j = -c^j c^i$, etc. The elements $c^i$ act on
the basis vectors according to:
\begin{equation}
\label{norm}
c^{i_1}\!\ldots c^{i_m}(T_{j_1}, \ldots, T_{j_m})=\det(\delta_j^i)\,,
\end{equation}
Any product of $n+1$ such elements must vanish by symmetry,
so the number of independent products of Grassman generators
is given by
\[
1 + n + \binom{n}{2} + \binom{n}{3} + \dots + \binom{n}{n-1} + \binom{n}{n}
= (1 + 1)^n
\]
We denote the algebra generated by these $2^n$ elements by $\F$,
and we remark that $\F$ has the structure of a fermionic Fock space,
with multiplication by the various $c^i$'s playing the role
of creation operators for different species of fermions.

In this notation, a general $n$-cochain takes the form
\begin{equation}\label{gen}
u = \frac{1}{n!}\,u_{i_1\ldots i_n}^{\alpha}c^{i_1}\!\ldots c^{i_n}e_\alpha\,,
\end{equation}
and the coboundary operator $\delta$ assumes the BRST-like form
\begin{equation}\label{min}
\delta = c^i t_i - \half {f_{ij}}^{k}\,c^i c^j \f{\d}{\d c^k}\ .
\end{equation}
As an odd differential operator, $\delta$ acts upon
each $c$ and $e$ in (\ref{gen}) by
\begin{equation}
\label{act}
\delta e_\alpha = c^i\, {(t_i)_\alpha}^\beta e_\beta\,, \ \ \ \ \
\delta c^i = - \half \, {f^i}_{jk}\,c^j c^k\,, \ \ \ \ \
\delta\circ c^i=\delta c^i - c^i\circ\delta\,.
\end{equation}
Viewing $c^i \in \g^*$ as left invariant forms subject to the
Maurer-Cartan equation, Eq.~(\ref{min}) becomes
\begin{equation}
\label{diff}
\delta = c^i t_i + d\,, \ \ \ \ \ \ \
d c^i = -\frac{1}{2}\,f_{jk}^{i}\,c^j c^k\,,
\end{equation}
and we deduce $\delta^2=0$ very easily:
\begin{equation}
\label{}
(c^i t_i + d)^2 = d^2 + t_i\,(d\circ c^i + c^i d) + c^j c^k t_j t_k
= t_i\,d c^i + \frac{1}{2}\,c^j c^k\,[t_j, t_k] = 0\,.
\end{equation}
The relations \eqref{diff} show that the BRST differential is a
homological perturbation of the differential of the Koszul-Tate complex.

\section{Faddeev-Popov Quantization}

Consider a Yang-Mills theory,
take $\g$ to be the Lie algebra of the gauge group, and recall that
$t_i$ denote generators of $\g$.
In this context, the operator (\ref{min}) is known as the minimal nilpotent
extension of $c^i t_i$. Other nilpotent extensions are
possible; a conventional form of the BRST operator in gauge theory is:
\begin{equation}
\label{BRST}
\delta =
c^i t_i - \half {f_{ij}}^k \,c^i c^j \f{\d}{\d c^k}
+ b_i \f{\d}{\d \bar{c}_i}\,.
\end{equation}
where $t_i$ are generators of gauge transformations of original
(matter and gauge) fields $\varphi$ present in the classical
theory, $c^i$ and $\bar{c}_i$ are ghosts and antighosts, and $b_i$
are auxiliary boson fields. We treat $c^i$ and $\bar{c}_i$ as
independent fermionic fields, of ghost number $+1$ and $-1$
respectively. The operator (\ref{BRST}) is nilpotent, odd, of
ghost number +1, and acts on the fields giving their variations
under BRST transformations:
\begin{align}
\delta\varphi &= c^i t_i \varphi \label{trans} \\
\delta c^i &= -\frac{1}{2}f_{jk}^{i}\,c^j c^k \\
\delta\bar{c}_i &= b_i \\
\delta b_i &= 0
\end{align}
An object is called \emph{BRST invariant} when it is annihilated by
$\delta$\,. In the BRST gauge fixing procedure, the gauge-fixed
Lagrangian has to be chosen in the form
\begin{equation}
\label{Lagr}
L = L_{\text{inv}} + \delta\Lambda
\end{equation}
with $L_{\text{inv}}(\varphi)$ invariant (and
$\Lambda(\varphi,c,\bar{c},b)$ non-invariant) under (\ref{trans})\,,
i.e., under local gauge transformations with ghosts $c^i$ playing
the role of gauge parameters. Also, the gauge fixing term
$\Lambda$ should be of ghost number -1. Due to $\delta^2 = 0$\,, gauge
invariance of $L_{\text{inv}}$ means BRST invariance of the total
Lagrangian $L$\,, and vice versa. For instance, the Faddeev-Popov
Ansatz is generally recovered by the choice
\begin{equation}
\label{FP}
\Lambda = \Phi^i\bar{c}_i \ \ \ \ \ \Rightarrow \ \ \ \ \
L = L_{\text{inv}} + c^j t_j\Phi^i\bar{c}_i + \Phi^i b_i
\end{equation}
where $\Phi^i(\varphi)$ are gauge-fixing conditions. Subsequent
functional integration over $b$ yields the delta-function of
$\Phi$ (fixing the gauge)\,, and functional integration over
$c,\,\bar{c}$ yields the determinant $\text{det}(t_j\Phi^i)$ of
variations under the gauge transformations, introduced by Faddeev
and Popov.

We now discuss the relationship of physical operators and
states to the BRST invariant objects.
First, consider a functional integral of the type
\begin{equation}
\label{Zdelta}
Z_\delta = \int\,\mathcal{D}\psi\,Y\,\delta X\,e^{iL}\,,
\end{equation}
where $\psi$ indexes all fields in the theory (including ghosts),
$Y$ is BRST invariant and $X$ arbitrary. Then, assuming BRST invariance
of functional measure (no anomalies), we see that
\begin{equation}
\label{}
Z_\delta = \int\,\delta(\mathcal{D}\psi\,Y\,X\,e^{iL}) = 0\,,
\end{equation}
as a (zero) result of a mere change of integration variables. But this
simple fact immediately implies
that amplitudes which we would like to call physical,
\begin{equation}
\label{phys}
Z_{\text{ph}}=
\int\,\mathcal{D}\psi\,Y e^{i(L_{\text{inv}}+\delta\Lambda)}\,,
\end{equation}
do not depend on the choice of gauge fixing term $\Lambda$\,, because
any change in $\Lambda$ produces additional contributions of the form
(\ref{Zdelta})\,.

Thus physical operators (or states) are classes of objects,
of the form $Y+\delta X$\,, with $Y$ being BRST invariant,
$\delta Y = 0$\,. Physical objects are associated
with cohomology classes of the BRST operator $\delta$\,. Objects of the
type $\delta X$  (those in zero cohomology) are called spurious.
As a rule, physical field $Y$ is also required to be of definite
ghost number (usually, 0)\,.

\section{Lie Algebra Deformations and the BRST
Operator}

The operator
\begin{equation} \label{defQ}
Q = c^i t_i - \half {f_{ij}}^k \,c^i c^j \f{\d}{\d c^k} \,.
\end{equation}
depends on the Lie algebra $\g$ through its structure constants
${f_{ij}}^k$, and on a $\g$-module $V$ in which the elements of
some basis for $\g$ act via representation matrices $t_i$. Thus
$Q$ is an algebraic object which is completely determined by an
ordered pair $(\g, V)$ where $\g$ is a finite-dimensional Lie
algebra and $V$ is a $\g$-module.

Let $\g(t)$ denote a deformation of the type considered in Section
\ref{sec:deformations}. We propose that invariants of deformations
$\g(t)$ can be constructed using the operators \eqref{defQ}.
Obvious invariants such as the trace do not yield interesting
results, since the BRST operator \eqref{defQ} is traceless in a
large class of examples. However, more sophisticated invariants
inspired by quantum field theory, such as heat kernel-regularized
traces of combinations of $Q$ with other operators, are known to
exist. A detailed study of one class of such invariants is due to
Jaffe \cite{QHA,QI}, and we consider the application of Jaffe's
theory to Lie algebras in Section \ref{sec:jaffe}.

The scheme $\M_n$ discussed in Section \ref{sec:deformations} is,
as one would expect, quite a complicated object; by definition it
contains all Lie algebras of dimension $n$. One way of elucidating
some of its structure is to identify path-components using
invariants from quantum field theory.

\section{Quantum Invariants} \label{sec:jaffe}

Let $\H$ be a Hilbert space, with an operator $\gamma$ on $\H$
that is both self-adjoint and unitary. Such an operator is called
a $\Z_2$-grading, because $\H$ splits into
$\pm 1$ eigenspaces for $\gamma$, i.e. $\H = \H_+ \oplus \H_-$.
Let $X = \{a_0, \ldots, a_n\}$ be a
set of operators on $\H$ which we will call \emph{vertices}.
The \emph{heat-kernel regularization density} of this set of vertices
is defined to be
\[
X(s) = \begin{cases} b_0 e^{-s_0 Q^2} b_1 e^{-s_1 Q^2} \ldots b_n e^{-s_n Q^2},
& \te{ when every } s_j > 0 \\
0, & \te{ otherwise}
\end{cases}
\]
and the Radon transform is $\hat X(\beta) = \int X(s) [d^n
s]_\beta$, where $[d^n s]_\beta$ is Lebesgue measure on the
hyperplane $\sum s_j = \beta$. Let $U(g)$ be a continuous unitary
representation of a compact Lie group $G$ on $\H$. The expectation
of a heat-kernel regularization $\hat X$ is defined by
\[
\<\hat X; g\> = \Tr(\gamma U(g) \hat X)
\]
$Q$ is a self-adjoint operator commuting with $U(g)$, called the
\emph{supercharge}, which is odd with respect to $\gamma$ and we
assume that $e^{-\beta Q^2}$ is trace class $\forall \ \beta > 0$.

\begin{definition}
The JLO cochain is the expectation whose $n$th component is defined by
\[
\tau^{JLO}_n(a_0, \ldots, a_n; g) = \< a_0, da_1, \ldots, da_n; g\>
\]
\end{definition}
The most general structure necessary to define a JLO cocycle is a
\emph{$\Theta$-summable fractionally differentiable structure,}
which is defined as a sextuple
\[
\{\H, Q, \gamma, G, U(g), \A\},
\]
where $\H, Q, \gamma, G, U(g)$ are all as defined above, and $\A$
is an algebra of operators (actually a subalgebra of an
interpolation space; see \cite{QHA}) which is pointwise invariant
under $\gamma$ and for which $U(g) \A U(g)^* \subset \A$. Further
references to ``cyclic cohomology'' refer to cohomology of the
algebra $\A$.

The JLO expectation is a cocycle in the sense of entire cyclic
cohomology theory, its application to physics is that pairing an
operator with a family of JLO cocycles, each coming from a
fractionally-differentiable structure on $\H$, gives a natural
generalization of a well-known equivariant index \eqref{equiv-ind}
in supersymmetric physics. Moreover, algebraic properties of
cyclic cohomology imply that the equivariant index is a homotopy
invariant. The equivariant index is defined by
\begin{equation} \label{equiv-ind}
\fZ^{Q(\lambda)}(a; g) = \frac{1}{\sqrt{\pi}} \int_{-\infty}^\infty
e^{-t^2} \Tr(\gamma U(g) a \, e^{-Q(\lambda)^2 + i t \, da}) \, dt
\end{equation}
and it is shown in \cite{QHA} that
\begin{eqnarray*}
\fZ^{Q(\lambda)}(a; g) &=& \< \tau^{JLO}(\lambda), a\>  \\
&& \te{ and } \\
\frac{d}{d\lambda}\< \tau^{JLO}(\lambda), a\> &=& 0
\end{eqnarray*}
All known cocycles in entire cyclic cohomology are of the form $\tau^{JLO}$
for some choice of a $\Theta$-summable fractionally differentiable structure.

A simple proof that $\fZ^{Q(\lambda)}(a; g)$ is a geometric
invariant is given in \cite{QI}, however, the proof relies on
analytic hypotheses which have to be checked in each case of
interest. Following \cite{QI}, we note that the following
regularity hypotheses are sufficient to establish
$\fZ^{Q(\lambda)}(a; g)$ as an invariant.
\begin {itemize}
\ritem{1.}
The operator $Q$ is self-adjoint operator on $\H$, odd with respect to
$\gamma$, and $e^{-\beta Q^2}$ is trace class for all $\beta>0$.

\ritem{2.}
For $\la\in J$, where $J$ is an open interval on the real line,
the operator $Q(\la)$ can be expressed as a perturbation of $Q$ in the form
\begin{equation}
Q(\la)=Q+W(\la)
\label{11.2} \end{equation}
Each $W(\la)$ is a symmetric operator on the domain $\D=C^\infty(Q)$.

\ritem{3.} Let $\la$ lie in any compact subinterval $J' \subset J$.
The inequality
\begin{equation}
W(\la)^2 \le a Q^2 + bI\;,
\label{11.3} \end{equation}
holds as an inequality for forms on $\D\times\D$.  The
constants $a<1$ and $b<\infty$ are independent of $\la$ in the compact
set $J'$.

\ritem{4.} Let $R=(Q^2 + I)^{-1/2}$. The operator $Z(\la)=R W(\la) R$
is bounded uniformly for $\la\in J'$, and the difference quotient
\begin{equation}
\frac{Z(\la)-Z(\la')}{\la-\la'}
\label{11.4} \end{equation}
converges in norm to a limit as $\la'\to\la \in J'$.

\ritem{5.} The bilinear form $d_\la a$ satisfies the bound
\begin{equation}
\norm{ R^\alpha d_\la a R^\beta } < M\;,
\label{11.5} \end{equation}
with a constant $M$ independent of $\la$ for $\la\in J'$.
Here $\alpha,\beta$ are non-negative constants and $\alpha+\beta <1$.
\end{itemize}

\begin{corollary}[Ref.~\cite{QI}]
If a family $Q(\lambda)$ satisfies conditions 1-5, then the numerical quantity
\[
\fZ^{Q(\lambda)}(a; g) = \frac{1}{\sqrt{\pi}} \int_{-\infty}^\infty
e^{-t^2} \Tr(\gamma U(g) a \, e^{-Q(\lambda)^2 + i t \, d_\lambda a}) \, dt
\]
does not depend upon $\lambda$, and hence is constant on
path-components.
\end{corollary}

In the following theorem, we check the hypotheses 1-5 for the
BRST operator.

\begin{theorem} \label{thm:brstreg}
A family $Q(\lambda)$ of BRST operators formed by a continuous
deformation of Lie group structures satisfies
hypotheses 1-5. The numerical quantity
\[
\fZ^{Q(\lambda)}(a; g) = \frac{1}{\sqrt{\pi}} \int_{-\infty}^\infty
e^{-t^2} \Tr(\gamma U(g) a \, e^{-Q(\lambda)^2 + i t \, d_\lambda a}) \, dt
\]
is an invariant.
\end{theorem}

\proofthm{thm:brstreg} Since $Q(\la)^2 = 0$, $e^{-\beta Q^2}$ must
be trace class. If the grading $\gamma$ is by ghost number, then
to see that $Q$ is $\gamma$-odd, note that each term in $Q$
increases total ghost number by one. This proves {\bf 1}.

{\bf 2} is also easy, setting $W(\la) = Q(\la) - Q$ and $Q =
Q(0)$. Each $Q(\la)$ is symmetric, thus $W$ is also.

To prove {\bf 3}, we wish to show that there exist
$\la$-independent constant $b < \infty$ such that $bI + \{Q(\la),
Q\}$ is a positive operator for all $\lambda \in J'$. Let
$E_{\te{min}}(\cdot)$ denote the lowest eigenvalue of a
finite-dimensional matrix. $E_{\te{min}}(\{Q(\la), Q\})$ is a
continuous function of $\la$ over the compact interval $J'$, so it
is bounded below and above. We choose
\[
b > \left| \, \min_\la E_{\te{min}}(\{Q(\la), Q\}) \, \right|
\]

In property {\bf 4}, nilpotency requires $R = I$, and $Z(\la) =
W(\la)$. We then wish to show that
\[
(\la - \la')^{-1} (Q(\la) - Q(\la'))
\]
converges in norm to a limit as $\la' \to \la$ in $J'$.
\[
Q(\la) - Q(\la') = \half ({f(\la')_{ij}}^{k} - {f(\la)_{ij}}^{k})  \,c^i c^j \f{\d}{\d c^k}
\]
Therefore the difference quotient \eqref{11.4} certainly converges if
the structure constants ${f(\la)_{ij}}^k$ are differentiable in
$\la$.

The same assumption implies that $\lambda \to W(\la)$ is a
differentiable map, therefore $\la \to \| W(\la) \|$ is uniformly
bounded. The proof of {\bf 5} is similar. $\Box$

\vskip 0.1 in

To simplify calculations, for the moment we take $U(g)$ to be a
trivial representation. \footnote{it would be interesting to
incorporate $U(g)$-dependent terms, since the Lie group $G$ which
is the domain of the unitary representation $U$ can be different
from the simply connected group(s) $G(\lambda)$ which generate the
BRST transformation.} Since $Q$ is nilpotent, the invariant is
given by the integral of
\[
\pi(\lambda,t) \equiv \Tr(\gamma a  e^{-i t [Q(\lambda), a]} )
\]
with respect to a Gaussian measure in the variable $t$.

Let us compute $Q$ on a basis element of $\F \otimes V$. The
Maurer-Cartan equation for ghosts gives
\[
Q c^k = -\frac{1}{2} f_{ij}^k c^i c^j
\]
The product rule for the Grassmann derivative is
\begin{eqnarray*}
\f{\d}{\d c^k} (c_{i_1} c_{i_2} \ldots c_{i_r}) &=& \delta_{k i_1}
c_{i_2} \ldots c_{i_r} - \delta_{k i_2} c_{i_1} c_{i_3} \ldots
c_{i_r} + \dots \\
&& \hspace{1.5in} + (-1)^{r-1} \delta_{k i_r} c_{i_1} \ldots
c_{i_{r-1}} \\
&=& \sum_p (-1)^{p-1} \delta_{k\, i_p} c_{i_1} \ldots
\widehat{c}_{i_p} \ldots c_{i_r}
\end{eqnarray*}

We consider a homogeneous element of $\F \ox V$, which takes the
form $X = c^{i_1} \dots c^{i_n} \otimes v$ where $n$ is arbitrary.
We compute
\[
QX = (-1)^n c^{i_1} \dots c^{i_{n}} c^j \otimes t_j[v] + \Big(
f_{ij}^k c^i c^j \sum_p (-1)^{p-1} \delta_{k\, i_p} c^{i_1} \ldots
\widehat{c}^{i_p} \ldots c^{i_r} \Big) \otimes v
\]
This shows how to find the matrix elements of $Q$ in any basis of
$\F \otimes V$.

Although finite-dimensional, from a computational standpoint the
vector space $\F \ox V$ is usually rather large. In the case of
dimension 3 algebras (c.f. Example \ref{ex:nonisomorphic}), we
need three $c$'s, and the exterior algebra $\F$ will be $2^3 = 8$
dimensional. If $V$ is the adjoint representation, $\dim(\F \ox V)
= 24$.  In practice, one may choose the ``standard'' basis of $\F$
given by lexicographically ordered products of the $c_i$'s. On
this space, left multiplication by $c^j$ is a linear operation and
its matrix is identical to a permutation matrix up to factors of
$\pm 1$ in the various matrix elements. Similarly, $\d/\d c^i$ is
representable by a permutation matrix up to signs.

\begin{definition}
A \emph{monoid} is a semigroup with identity. Define ${\cal G}_N$
to be the set of all $N \times N$ matrices in which each row and
each column has at most one nonzero entry, and that entry is $\pm
1$. Then ${\cal G}$ is closed under products, but contains
non-invertible elements (those with a zero row or column), and so
it is a monoid.
\end{definition}

Left multiplication by $c^j$ is a linear operation on $\F$ with
non-trivial kernel. Its matrix is given by a permutation matrix,
except that there can be factors of $\pm 1$ in the various matrix
elements, and there can be rows or columns with all zeros; $\d/\d
c^i$ is representable by another such matrix.

\begin{lemma} \label{GLA}
Let $c^j$ ($j = 1, \ldots, n$) be ``Faddeev Popov Ghosts'' in the
sense considered above, let $A$ denote the $N = 2^n$ dimensional
algebra generated by $\{ c^j \}$ over the complex numbers, and let
$C^j$ denote the operator of left multiplication by $c^j$ on $A$.
Then the natural representation
\[
    \rho : R \To M_{N \times N}(\C)\, ,
\]
takes values in the monoid ${\cal G}_{N}$, where
\[
    R
    =
    \C \left[
    C^1, \ldots, C^n, \f{\d}{\d c^1}, \ldots, \f{\d}{\d c^n}
    \right] \, .
\]
\end{lemma}

Determination of the $\dim(\F \ox V)^2$-dimensional matrix of $Q$
is approachable by computer algebra methods. A number of commonly
available computer algebra systems contain facilities for working
with finitely generated algebras defined by generators and
relations, so the system facilitates the creation of subroutines
which represent the ghost algebra $A$ and the operators in the
ring $R$. In writing such a program, it is extremely useful to
keep Lemma \ref{GLA} in mind. In performing these calculations for
a 24-dimensional example, we discovered that all of the matrices
$Q(\lambda)$ for different values of $\lambda$ have the same
Jordan canonical form. We have not yet determined whether this
must always happen for differentiable families of BRST operators.


\end{document}